\documentclass[aps,prd,onecolumn,groupedaddress,showpacs,nofootinbib,amssymb]{revtex4}
\usepackage[dvips]{graphicx}
\usepackage{amssymb}
\usepackage{amsmath}
\usepackage{graphicx,,color}
\usepackage{amsfonts}
\usepackage{bm}
\usepackage{cancel}
\usepackage{comment}
\newcommand\e{\mathrm{e}}

\usepackage{babel}

\allowdisplaybreaks[4]

\begin{document}

\tolerance=5000

\title{2D $F(R)$ gravity and AdS$_2$/CFT$_1$ correspondence}
\author{S.~Nojiri,$^{1,2}$}
\email{nojiri@gravity.phys.nagoya-u.ac.jp}
\author{S.~D.~Odintsov,$^{3,4}$}
\email{odintsov@ice.cat}
\affiliation{$^{1)}$ Department of Physics, Nagoya University, Nagoya 464-8602, Japan \\
$^{2)}$ Kobayashi-Maskawa Institute for the Origin of Particles
and the Universe, Nagoya University, Nagoya 464-8602, Japan \\
$^{3)}$ ICREA, Passeig Luis Companys, 23, 08010 Barcelona, Spain \\
$^{4)}$ Institute of Space Sciences (ICE, CSIC) \\
C. Can Magrans
s/n, 08193 Barcelona, Spain \\
}

\begin{abstract}
We studied the canonical structure of 2D $F(R)$ gravity. 
Its equivalence with Jackiw-Teitelboim gravity is demonstrated when no matter presents.
Then, due to AdS$_2$/CFT$_1$ correspondence, 
such $F(R)$ gravity is equivalent to the Sachdev-Ye--Kitaev models. 
The singular $D\to 2$ limit of $F(R)$ gravity is also studied. 
It is shown that in such a limit AdS$_2$/CFT$_1$ correspondence is not realized.
\end{abstract}

%PACS numbers: 04.50.Kd, 95.36.+x, 98.80.-k, 98.80.Cq
%\pacs{04.50.Kd, 95.36.+x, 98.80.-k, 98.80.Cq,11.25.-w}

\maketitle

\noindent {\bf 1. Introduction}
For the resolution of the problem of the early- and late-time acceleration of the universe, several modified gravities have been developed
(for general review see \cite{Capozziello:2011et, Nojiri:2010wj}).
Among these models of modified gravity theories, the most realistic ones are different versions of $F(R)$ gravity,
which may even describe the whole universe's evolution from inflation to the dark energy epoch \cite{Nojiri:2003ft}.

  From another point, recently the wonderful equivalence between 2D Jackiw-Teitelboim (JT) gravity \cite{Teitelboim:1983ux, Jackiw:1984je}
and so-called the Sachdev-Ye--Kitaev (SYK) models have been discovered in Refs.~\cite{Sachdev:1992fk, Kitaev}.
In the present letter, we study the canonical structure of 2D $F(R)$ gravity and its relation to JT gravity.
It is shown that due to its equivalence with the JT gravity, the theory enjoys AdS$_2$/CFT$_1$ correspondence.
Its singular $D\to 2$ limit is also discussed.
It is demonstrated that gravity theory in such a limit does not have AdS$_2$ solutions.
As a result, in this case, there is no AdS$_2$/CFT$_1$ correspondence, and the relation with the SYK models is not realized.

Let us now show that $F(R)$ gravity is a unitary one and is equivalent to other 2D theories
like the dilaton gravity and the Jackiw-Teitelboim (JT) gravity when no matter is included.

The action of $F(R)$ gravity in $D$-dimensions,
\begin{align}
\label{FR1}
S=\frac{1}{2\kappa^2} \int dx^D \sqrt{-g} F(R)\, ,
\end{align}
can be rewritten as
\begin{align}
\label{FR2}
S=\frac{1}{2\kappa^2} \int dx^D \sqrt{-g} \left( \phi R - V(\phi) \right)\, .
\end{align}
In fact, by the variation of the action (\ref{FR2}) with respect to $\phi$, we find
$R=V'(\phi)$, which can be solved with respect to $\phi$, $\phi=\phi(R)$.
By substituting the expression into the action (\ref{FR2}), we obtain the action (\ref{FR1}) with $F(R) = \phi(R) R - V\left(\phi\left(R\right) \right)$.
The model (\ref{FR2}) is a kind of dilaton gravity, whose relations with the Sachdev-Ye--Kitaev (SYK) model have been well-studied \cite{Jensen:2016pah, Maldacena:2016upp}
in two dimensions.

By the variation of the action (\ref{FR2}) with respect to the metric, we obtain
\begin{align}
\label{FR5}
0= \phi \left( \frac{1}{2} g_{\mu\nu} R - R_{\mu\nu} \right) + \nabla_\mu \nabla_\nu \phi - g_{\mu\nu} \Box \phi
  - \frac{1}{2} g_{\mu\nu} V(\phi) \, .
\end{align}
If we consider the scale transformation $g_{\mu\nu}\to \e^\sigma g_{\mu\nu}$, the action (\ref{FR2}) is transformed as
\begin{align}
\label{WFR2}
S=\frac{1}{2\kappa^2} \int dx^D \sqrt{-g} \e^{\frac{D}{2}\sigma} \left\{ \phi \left(R - (D-2)\Box \sigma
  - \frac{(D-3)(D-2)}{4}\partial^\mu \sigma \partial_\mu \sigma \right)\e^{-\sigma} - V(\phi) \right\}\, .
\end{align}
Except for two-dimensional case, $D\neq 2$, by choosing $\sigma$ so that
$\e^{\frac{D-2}{2}\sigma}\phi=1$,
we obtain the well-known scalar-tensor theory,
\begin{align}
\label{WFR4}
S=\frac{1}{2\kappa^2} \int dx^D \sqrt{-g} \left\{ R - \frac{(D-3)(D-2)}{4}\partial^\mu \sigma \partial_\mu \sigma
  - \e^{\frac{D}{2}\sigma} V\left( \phi=\e^{- \frac{D-2}{2}\sigma} \right) \right\}\, .
\end{align}
Therefore, general scalar-tensor theory is equivalent to $F(R)$ gravity.

The scalar-tensor theory (\ref{WFR4}) shows that the model does not include any ghosts.
On the other hand, in two dimensions, $D=2$, the action (\ref{WFR2}) has the following form,
\begin{align}
\label{WFR5}
S=\frac{1}{2\kappa^2} \int dx^2 \sqrt{-g} \left\{ \phi R - \e^\sigma V(\phi) \right\}\, ,
\end{align}
and we cannot rewrite the action in the scalar-tensor form as in (\ref{WFR4}).
One can choose, however, $\sigma$ as $\e^\sigma V(\phi) = \Lambda$
with a constant $\Lambda$, which can be identified with a cosmological constant.
For the choice $\e^\sigma V(\phi) = \Lambda$, the action (\ref{WFR5}) reduces,
\begin{align}
\label{WFR7}
S=\frac{1}{2\kappa^2} \int dx^2 \sqrt{-g} \left\{ \phi R - \Lambda \right\}\, .
\end{align}
Instead of $\e^\sigma V(\phi) = \Lambda$, we may choose other form of $\sigma$,
as $\e^\sigma V(\phi) = \phi \Lambda$ with a constant $\Lambda$, again.
Under the choice $\e^\sigma V(\phi) = \phi \Lambda$, the action (\ref{WFR5}) reduces,
\begin{align}
\label{WFR7B}
S=\frac{1}{2\kappa^2} \int dx^2 \sqrt{-g} \phi \left\{ R - \Lambda \right\}\, ,
\end{align}
which is nothing but the action of the JT gravity \cite{Teitelboim:1983ux, Jackiw:1984je}.
Therefore, $F(R)$ gravity without matter in two dimensions is equivalent to the Jackiw-Teitelboim gravity.
More specific choice could be given by $\sqrt{-g} \e^\sigma V(\phi) = L^2$ with a constant $L$.

We have shown the equivalence between $F(R)$ gravity and other gravity theories in two dimensions on the classical level.
If we consider the quantum theory by using the path integral formalism, we need to define the integration measures for the fields.

\

\noindent {\bf 2. Canonical Structure of 2D $F(R)$ gravity}
In two dimensions, one can choose the conformal gauge condition,
$g_{\mu\nu}=\e^{2\gamma} \eta_{\mu\nu}$,
where $\eta_{\mu\nu}$ is the metric in the flat two-dimensional space-time
$\left( \eta_{\mu\nu} \right) = \left( \begin{array}{rr} -1 & 0 \\ 0 & 1 \end{array} \right)$.
Then $(0,0)$, $(1,1)$, and $(0,1)$ components of Eq.~(\ref{FR5}) are
\begin{align}
\label{FR9}
0 = \phi_{,11} - \gamma_{,0} \phi_{,0} - \gamma_{,1} \phi_{,1} + \frac{\e^{2\gamma}}{2} V(\phi) \, , \
0 = \phi_{,00} - \gamma_{,0} \phi_{,0} - \gamma_{,1} \phi_{,1} - \frac{\e^{2\gamma}}{2} V(\phi) \, , \
0 = \phi_{,01} - \gamma_{,0} \phi_{,1} - \gamma_{,1} \phi_{,0} \, ,
\end{align}
and the equation $R=V'(\phi)$ for the action~(\ref{FR2}) has the following form
\begin{align}
\label{FR10}
0 = 2 \left( - \gamma_{,00} + \gamma_{,11} \right) + \e^{2\gamma}V'(\phi)\, .
\end{align}
Under the conformal gauge condition $g_{\mu\nu}=\e^{2\gamma} \eta_{\mu\nu}$, one may write the action (\ref{FR2}) which is completely equivalent to the action of the $F(R)$ gravity,
\begin{align}
\label{FR11}
S= \frac{1}{2\kappa^2} \int dx^2 \left( - 2 \phi \eta^{\rho\sigma} \gamma_{,\rho\sigma} - \e^{2\gamma} V(\phi) \right)
= \frac{1}{2\kappa^2} \int dx^2 \left( 2 \eta^{\rho\sigma} \phi_{,\rho} \gamma_{,\sigma} - \e^{2\gamma} V(\phi) \right) \,.
\end{align}
The variation of the action with respect to $\gamma$ gives a combination of the first two equations in (\ref{FR9}), that is, the trace part,
$0= \phi_{,00} - \phi_{,11} - \e^{2\gamma} V(\phi)$.
The combination of the first two equations in (\ref{FR9}) independent of the above trace equation is given by
$0= \phi_{,00} + \phi_{,11} - 2 \gamma_{,0} \phi_{,0} - 2 \gamma_{,1} \phi_{,1}$.
The combination of the kinetic term $ \phi \eta^{\rho\sigma} \gamma_{,\rho\sigma}$ in the action (\ref{FR11}) could indicate that there appears a ghost,
which may be special in two dimensions.

To investigate if there could be a ghost or not, we use the Hamiltonian formalism.
The action (\ref{FR11}) shows that the conjugate momenta $\pi_\phi$ and $\pi_\gamma$ which are conjugate to $\phi$ and $\gamma$ are given by
$\pi_\phi = - \frac{1}{\kappa^2}\gamma_{,0}$ and $\pi_\gamma= - \frac{1}{\kappa^2}\phi_{,0}$, respectively, and the Hamiltonian is
\begin{align}
\label{FR15}
H = \int dx^1 \left\{ - \kappa^2 \pi_\phi \pi_\gamma - \frac{1}{\kappa^2} \left( \phi_{,1} \gamma_{,1} -\frac{1}{2} \e^{2\gamma} V(\phi) \right) \right\} \, .
\end{align}
The Hamiltonian density is the first equation in (\ref{FR9}) times $\frac{1}{\kappa^2}$ and the total derivative term
\begin{align}
\label{Hd} \mathcal{H} \equiv - \kappa^2 \pi_\phi \pi_\gamma - \frac{1}{\kappa^2} \left( \phi_{,1} \gamma_{,1} -\frac{1}{2} \e^{2\gamma} V(\phi) \right)
= \frac{1}{\kappa^2} \left( - \gamma_{,0} \phi_{,0} - \gamma_{,1} \phi_{,1} + \frac{1}{2}\e^{2\gamma} V(\phi) \right) + \partial_1 \left( \frac{\phi_{,1}}{\kappa^2} \right) \, .
\end{align}
Therefore the Hamiltonian constraint is not given by $\mathcal{H}=0$ but by the first equation (\ref{FR9}).

The first and last equations in (\ref{FR9}) can be regarded with the constraint equations
\begin{align}
\label{FR16}
0 = C_{(1)} \equiv \phi_{,11} - \kappa^4 \pi_\phi \pi_\gamma - \gamma_{,1} \phi_{,1} + \frac{\e^{2\gamma}}{2} V(\phi) \, , \quad
0= C_{(2)} \equiv - \pi_{\gamma,1} + \pi_\phi \phi_{,1} + \gamma_{,1} \pi_\gamma \, .
\end{align}
In order to find the secondary constraints, we define the Poisson bracket for two physical quantities $F\left( x_{(1)} \right)$ and $G\left( x_{(2)} \right)$ as follows,
\begin{align}
\label{FR18}
& \left[ F\left( x_{(1)} \right), G\left( x_{(2)} \right) \right] \nonumber \\
& \quad \equiv \int dx^1 \left\{ \frac{\partial F\left( x_{(1)} \right)}{\partial \phi(x)}\frac {\partial G\left( x_{(2)} \right)}{\partial \pi_\phi(x)}
  - \frac{\partial F\left( x_{(1)} \right)}{\partial \pi_\phi(x)}\frac {\partial G\left( x_{(2)} \right)}{\partial \phi(x)}
+ \frac{\partial F\left( x_{(1)} \right)}{\partial \gamma(x)}\frac {\partial G\left( x_{(2)} \right)}{\partial \pi_\gamma(x)}
  - \frac{\partial F\left( x_{(1)} \right)}{\partial \pi_\gamma(x)}\frac {\partial G\left( x_{(2)} \right)}{\partial \gamma(x)} \right\} \, .
\end{align}
By using the Poisson bracket, we find
\begin{align}
\label{FR18BB}
\left[ C_{(1)} \left(x_{(1)}\right), H \right] = \kappa^2 C_{(2),1} \left( x_{(1)} \right) \, , \quad
\left[ C_{(2)} \left(x_{(1)}\right), H \right] = \frac{1}{\kappa^2} C_{(1),1} \left( x_{(1)} \right) \, .
\end{align}
Therefore if $C_{(1)}=C_{(2)}=0$, we obtain $\left[ C_{(1)} \left(x_{(1)}\right), H \right] = \left[ C_{(2)} \left(x_{(1)}\right), H \right] = 0$ and there does not appear any secondary constraint.
Hence we have four canonical variables $\phi$, $\pi_\phi$, $\gamma$, and $\pi_\gamma$ and two constraints, we have two independent canonical variables.
This shows that although the action (\ref{FR11}) seems to generate a ghost, the constraints might delete the ghost.

\

\noindent {\bf 3. AdS$_2$/CFT$_1$ correspondence}
We now discuss the correspondence between 2D anti-de Sitter gravity and the conformal field theory in one dimension, that is, the AdS$_2$/CFT$_1$ correspondence.

The conformal field theory in one dimension is nothing but quantum mechanics with conformal or scale invariance.
A candidate of the conformal quantum mechanics is given by the Sachdev-Ye--Kitaev (SYK) model \cite{Sachdev:1992fk, Kitaev},
which is given by $2N$ Majorana fermions $\psi^a$, $a=1,2, \cdots, 2N$ and the Hamiltonian is given by
\begin{align}
\label{SYK1}
H=\sum_{a,b,c,d=1}^{2N} \frac{J_{abcd}}{4!}\psi^a \psi^b \psi^c \psi^d\, .
\end{align}
Here $J_{abcd}$ is a random source whose average vanishes, $\overline{J_{abcd}}=0$, and the average of the square of $J_{abcd}$ is given
by $\sum_{a,b,c,d=1}^{2N} \overline{\left(J_{abcd}\right)^2} = \frac{3! J^2}{\left( 2N \right)^3}$ with a constant $J$.
The AdS$_2$/CFT$_1$ correspondence of this model has been well-studied \cite{Jensen:2016pah,Maldacena:2016upp}

In the model, there appears a conformal symmetry at low energy.
Interestingly, this model saturates the chaos bound \cite{Maldacena:2015waa, Shenker:2013pqa}.
The conformal symmetry in $d$ dimensions is identical with the isometry of the AdS space-time in $D=d+1$ dimensions.
In this section, we check those 2D gravity models have AdS space-time as a solution.

In the case of $F(R)$ gravity (\ref{FR1}), the variation over the metric gives,
\begin{align}
\label{SYK2}
0= \frac{1}{2}g_{\mu\nu} F(R) - R_{\mu\nu}F_R(R) + \nabla_\mu \nabla_\nu F_R(R)
  - g_{\mu\nu}\nabla^2 F_R (R) \, .
\end{align}
Here $F_R(R)\equiv \frac{dF(R)}{dR}$.
Because AdS space-time is the Einstein manifold, where the Ricci tensor $R_{\mu\nu}$ is covariantly constant, $\nabla_\rho R_{\mu\nu}=0$,
we may assume $R_{\mu\nu}=\frac{1}{2} R_0 g_{\mu\nu}$ in two dimensions.
Here $R_0$ is a constant corresponding to the scalar curvature.
Then Eq.~(\ref{SYK2}) reduces to the form, $0=F\left( R_0 \right) - R_0 F\left( R_0 \right)$, which is an algebraic equation for $R_0$.
Therefore if this algebraic equation has a negative solution for $R_0$, the anti-de Sitter space-time is a solution of the $F(R)$ gravity (\ref{FR1}) in two dimensions.

In the form of Eq.~(\ref{FR2}), the existence of the anti-de Sitter space-time as a solution and also
the AdS$_2$/CFT$_1$ correspondence of this model has been investigated in \cite{Jensen:2016pah}.
By assuming $R_{\mu\nu}=\frac{1}{2} R_0 g_{\mu\nu}$ and $\phi=\phi_0$ with constants $R_0$ and $\phi_0$,
The equation $R=V'(\phi)$ for the action~(\ref{FR2}) and Eq.~(\ref{FR5}) have the forms, $R_0=V' \left( \phi_0 \right)$ and $V \left( \phi_0 \right) = 0$.
Because $F(R) = \phi(R) R - V\left(\phi\left(R\right) \right)$, we find
\begin{align}
\label{SYK5}
F_R(R) = \left( R - V'\left( \phi \right) \right) \frac{d\phi}{dR} + \phi(R) = \phi(R)\, ,
\end{align}
where we used the equation $R=V'(\phi)$ for the action~(\ref{FR2}).
Therefore by using the equation $F(R) = \phi(R) R - V\left(\phi\left(R\right) \right)$ and (\ref{SYK5}), the condition $ V \left( \phi_0 \right) = 0$
can be rewritten as
\begin{align}
\label{SYK6}
0= V \left( \phi_0 \right) = \phi \left( R_0 \right) R_0 - F\left( R_0 \right) = F_R \left( R_0 \right) R_0 - F\left( R_0 \right) \, ,
\end{align}
which is nothing but the condition $0=F\left( R_0 \right) - R_0 F\left( R_0 \right)$ as expected.

As AdS$_2$/CFT$_1$ correspondence between the gravity model (\ref{FR2}) and the SYK model (\ref{SYK1}) has been established in \cite{Jensen:2016pah},
the equivalence between the model in (\ref{FR2}) and the $F(R)$ gravity model (\ref{FR1}) proves that the $F(R)$ gravity model (\ref{FR1})
in two dimensions is dual to the SYK model (\ref{SYK1}).

\

\noindent {\bf 4. Singular $D\to 2$ limit in $F(R)$ gravity}
In two dimensions, the Hilbert-Einstein action is a total derivative and the action does not contribute to any dynamics.
There was, however, a proposal that Einstein's theory in the limit of $D\to 2$ with a singular rescaling of
the gravitational coupling constant as $1/(D-2)$ \cite{Mann:1992ar} (see also \cite{Nojiri:2020tph}) may lead to non-trivial gravitational dynamics.
In this section, we consider the $D\to 2$ limit in the scalar-tensor theory (\ref{WFR4}).

By the variation of the action (\ref{WFR4}) with respect to the metric, we obtain
\begin{align}
\label{D2l1}
0= \frac{1}{2\kappa^2} \left[ \frac{1}{2} g_{\mu\nu} \left\{ R - \frac{(D-3)(D-2)}{4}\partial^\rho \sigma \partial_\rho \sigma
  - \e^{\frac{D}{2}\sigma} V\left( \phi=\e^{- \frac{D-2}{2}\sigma} \right) \right\}
  - R_{\mu\nu} + \frac{(D-3)(D-2)}{4}\partial_\mu \sigma \partial_\nu \sigma \right] \, ,
\end{align}
and by the variation with respect to $\sigma$, we find
\begin{align}
\label{D2l2}
0 = \frac{(D-3)(D-2)}{2}\nabla^2 \sigma - \frac{D}{2} \e^{\frac{D}{2}\sigma} V\left( \phi=\e^{- \frac{D-2}{2}\sigma} \right)
+ \frac{D-2}{2} \e^{\frac{D}{2}\sigma} V'\left( \phi=\e^{- \frac{D-2}{2}\sigma} \right) \, .
\end{align}
By multiplying (\ref{D2l1}) with $g^{\mu\nu}$, we obtain
\begin{align}
\label{D2l3}
0= \frac{1}{2\kappa^2} \left[ \frac{D-2}{2} \left\{ R - \frac{(D-3)(D-2)}{4}\partial^\rho \sigma \partial_\rho \sigma \right\}
  - \frac{D}{2} \e^{\frac{D}{2}\sigma} V\left( \phi=\e^{- \frac{D-2}{2}\sigma} \right) \right] \, .
\end{align}
Therefore in the limit of $D\to 2$, one gets $0= \e^\sigma V\left( \phi=1 \right)$,
which, gives $V\left( \phi=1 \right) = 0$ and it conflict for general $V\left( \phi \right)$.

There are several limits to obtaining non-trivial solutions.
An example is given by the limit $D\to 2$ after redefining $\kappa^2={\tilde \kappa}^2 \left(D - 2 \right)$ and
$\sigma = \tilde\sigma + \frac{2}{D}\ln \left( D -2 \right)$ and keeping $\tilde\kappa$ and $\tilde\sigma$ finite.
In the limit, Eqs.~(\ref{D2l3}) and (\ref{D2l2}) have the following non-trivial forms,
\begin{align}
\label{D2l5}
0= \frac{1}{2{\tilde\kappa}^2} \left[ \frac{1}{2} R - \e^{\tilde\sigma} V\left( \phi=1 \right) \right] \, , \quad
0 = - \frac{1}{2}\nabla^2 {\tilde\sigma} - \e^{\tilde\sigma} V\left( \phi=1 \right) \, ,
\end{align}
which could give non-trivial solutions.

We now consider the solution of the equations in Eq.~(\ref{D2l5}).
In the conformal gauge $g_{\mu\nu}=\e^{2\gamma} \eta_{\mu\nu}$, we find that
The equations in (\ref{D2l5}) have the following forms,
\begin{align}
\label{D2l7}
0= \gamma_{,00} - \gamma_{,11} - \e^{{\tilde\sigma}+2\gamma} V_0 \, , \quad
0 = \frac{1}{2}\left( {\tilde\sigma}_{,00} - {\tilde\sigma}_{,11} \right) - \e^{{\tilde\sigma} + 2\gamma} V_0 \, ,
\end{align}
Here $V_0\equiv V\left( \phi=1 \right)$.
By combining the above equations in (\ref{D2l7}), we obtain
\begin{align}
\label{D2l9B}
0= \gamma_{,00} - \gamma_{,11} - \frac{1}{2}\left( {\tilde\sigma}_{,00} - {\tilde\sigma}_{,11} \right)\, , \quad
0 = \frac{1}{2}\left( \rho_{,00} - \rho_{,11} \right) - 2 V_0 \e^\rho \, .
\end{align}
Here $\rho\equiv {\tilde\sigma} + 2\gamma$.
We should note that the second equation of (\ref{D2l9B}) is nothing but the Liouville equation, whose solutions are well-known.
As an example, we review the static case, where we find $0 = \frac{d}{dx} \left( - \frac{1}{4} {\rho_{,1}}^2 - 2 V_0 \e^\rho \right)$
and therefore $\frac{1}{4} {\rho_{,1}}^2 + 2 V_0 \e^\rho =C$ $\left(C :\ \mbox{constant}\right)$,
that is, $x = - \frac{4}{\sqrt{C}} \left( \theta - \theta_0 \right)$ $\left( \e^{-\frac{\rho}{2}}= \sqrt{\frac{2V_0}{C}} \cosh \theta \right)$.
Here $\theta_0$ is a constant of the integration.
Therefore we find $\e^{-\rho}= \frac{2V_0}{C} \cosh^2 \left( \frac{\sqrt{C}}{4} \left( x - x_0 \right) \right)$.
Here $x_0 \equiv \frac{4\theta_0}{\sqrt{C}}$.
The general solution of the first equation in (\ref{D2l9B}) when the solution does not depend on $t$ is
${\tilde\sigma} = 2 \gamma - 4 \gamma_0 - 4\gamma_1 x$,
where $\gamma_0$ and $\gamma_1$ are constants and the factor $-4$ is just for convenience.
Therefore we find $\rho=4\left( \gamma - \gamma_0 - \gamma_1 x \right) $, which tells
$\e^{2\gamma} = \frac{C\e^{ \gamma_0 + \gamma_1 x}}{2V_0 \cosh \left( \frac{\sqrt{C}}{4} \left( x - x_0 \right) \right)}$.
The conformal gauge condition $g_{\mu\nu}=\e^{2\gamma} \eta_{\mu\nu}$
shows that the metric is given by
\begin{align}
\label{D2l16}
ds^2 = \frac{C\e^{ \gamma_0 + \gamma_1 x}}{2V_0 \cosh \left( \frac{\sqrt{C}}{4} \left( x - x_0 \right) \right)} \left( - dt^2 + dx^2 \right) \, .
\end{align}
Even if we consider the adjustments of the parameters $\gamma_0$, $\gamma_1$, $V_0$, $C$, and $x_0$, or redefinition of the coordinates $t$ and $x$,
we cannot obtain the AdS$_2$ space-time.
In the AdS$_2$ space-time, the scalar curvature $R$ is constant.
In order that $R$ is constant, the equation $0= \e^\sigma V\left( \phi=1 \right) $ shows that $\tilde\sigma$ must be constant but
The first equation in (\ref{D2l5}) does not allow $\tilde\sigma$ to be constant.
Therefore the AdS$_2$ space-time cannot be a solution of the $D\to 2$ model even asymptotically.

To be sure, we may consider the modification of the $D\to 2$ limit.
For example, if we redefine $\tilde \sigma = \epsilon \eta - \ln \epsilon$ and $V\left( \phi=1 \right) = \epsilon U_0$,
with constants $\epsilon$ and $U_0$, in the limit of $\epsilon\to 0$,
the first equation in Eq.~(\ref{D2l5}) reduces to the form, $0= \frac{1}{2{\tilde\kappa}^2} \left[ \frac{1}{2} R - U_0 \right]$.
This equation seems to tell that the scalar curvature $R$ is a constant
$R=2 U_0$ and therefore the AdS space-time could be a solution.
In the limit of $\epsilon\to 0$, however, the second equation in (\ref{D2l5}) gives $0= U_0$
and therefore the scalar curvature $R$ vanishes, $R\to 0$.

This demonstrates that AdS space-time could not be a solution of the equations in (\ref{D2l5}) in any limit even asymptotically.
Therefore this model could not be related to the SYK model and the perspectives of the AdS$_2$/CFT$_1$ correspondence are not clear
in the singular $D\to 2$ gravity model here.

In summary, we have shown that $F(R)$ gravity in two dimensions is
equivalent to several  gravity theories like JT gravity if we neglect the
matter.
The equivalence between  JT gravity and the SYK models, which have conformal symmetry in one dimension,
has been well discussed from the viewpoint of the AdS$_2$/CFT$_1$
correspondence.
Therefore that $F(R)$ gravity in two dimensions is also equivalent to the
SYK models.

We also discussed the singular $D\to 2$ limit in the $F(R)$ gravity and we have shown that such a limit does not have AdS$_2$ solutions
and therefore there is no AdS$_2$/CFT$_1$ correspondence in this case.
However, the account of matter or quantum effects may qualitatively change
this picture.

\ 

\noindent
{\bf Acknowledgments}
This work is supported in part by MINECO (Spain), project PID2019-104397GB-I00 (SDO).

\end{document}